\documentclass[12pt,a4paper]{article}
\usepackage{graphicx}
\usepackage{amsmath}
\usepackage{siunitx}
\usepackage{fixltx2e}
\usepackage{hyperref}
\usepackage{pdflscape}

\usepackage{amssymb}

\begin{document}

\title{Study of the terahertz spectra of crystalline materials using NDDO semi-empirical methods: polyethylene, poly(vinylidene fluoride) form II and $\alpha$-D-glucose}

\author{P. Chamorro-Posada\\
  Dpto. de Teor\'{\i}a de la Se\~nal y Comunicaciones\\
 e Ingenier\'{\i}a Telem\'atica,\\
 Universidad de Valladolid, ETSI Telecomunicaci\'on,\\
 Paseo Bel\'en 15, 47011 Valladolid, Spain}

\maketitle
\begin{abstract}

Semi-empirical quantum chemistry methods offer a very interesting compromise between accuracy and computational load.  In order to assess the performance of NDDO methods in the interpretation of terahertz spectra, the low frequency vibration modes of three crystalline materials, namely,  polyethylene, poly(vinylidene fluoride) form II and $\alpha$-D-glucose have been studied using the PM6 and PM7 Hamiltonians and the results have been compared with the experimental data and former calculations.  The results show good qualitative or semi-quantitative agreement with the experimentally observed terahertz spectra.  

\end{abstract}

\section{Introduction}

Terahertz spectroscopy is an emerging characterization technique \cite{peiponen}.  The contributions from long range interactions of macro- and supra- molecular origin to the THz spectra provide very useful information on the morphology of materials.  Temperature-dependent lattice mode shifts, for instance, reveal information on the interactions between the amorphous and the crystalline phase of highly-crystalline polymers \cite{wietzke}.  THz spectra have also been shown to permit the identification of the transition from a 2D form to a 3D arrangement in a carbon nitride polymer \cite{chamorro1}.  The absorption in the THz has been demonstrated to display a direct correlation with the stacking order along the c axis in different types of carbon materials \cite{chamorro2}.

The complexity of the underlying physical effects affecting the observed attenuation, with contributions from intra- and inter-molecular modes and a strong impact of disorder, hinders the interpretation of the data.  Theoretical methods are essential for this task, and the versatility of computational quantum chemistry methods is particularly attractive for the discernment of THz spectra.

The theoretical prediction of the terahertz spectra of crystalline materials has been based mostly on force-field or density functional theory (DFT) methods \cite{peiponen}.  Ab-initio and density functional theory (DFT) methods are regarded as highly accurate in the prediction of molecular vibrations \cite{scott} when used with correction factors to compensate systematic deviations. Nevertheless, the relevance of long-range and intermolecular contributions can make the study of the THz vibrational spectra of materials very demanding in terms of the required solid-state calculations over large systems.  This imposes a severe limitation in the use of computationally costly ab-initio or DFT methods.  On the other hand, semi-empirical methods \cite{coolidge,fekete}, specially with the recently developed parallel implementations for shared-memory multiprocessor and massively-parallel graphics processing units architectures \cite{maia}, permit to study the vibrations in large molecular systems and crystals at reduced computation times.

Semi-empirical NDDO (neglect of the diatomic differential overlap) computations are based on approximations and data obtained from given training sets which calls for their assessment when employed in a given task, and can be placed at an intermediate theory level between force-field and ab-initio or DFT methods.  Poor performance of the AM1 \cite{AM1} and the PM3 \cite{PM3} methods reported in \cite{coolidge} disfavored their use in the analysis of vibrational spectra until the appearance of the PM6 parametrization \cite{PM6}, with improved accuracy in the prediction of geometries and the description of the material vibrations \cite{fekete}.  The recently developed PM7 \cite{PM7} method yields higher accuracy in the guessed structures, particularly for large molecular crystals, even though its preciseness in the description of vibration modes has not been systematically assessed.   In this work, the use of PM6 and PM7 in the study of the THz spectra of materials is addressed.  These methods have been shown to be very useful in the explanation of the observed THz spectra of various carbon materials \cite{chamorro2} and carbon nitride polymers \cite{chamorro1} but the accuracy of the methods must be contrasted by checking the results obtained for materials for which the terahertz properties are well-known.  In this work, we focus on three test cases that correspond to two crystalline polymers and a molecular crystal whose properties have been largely studied in this spectral region both experimentally and theoretically.

\section{Methods}

The computations have been performed using the MOPAC2012 software package \cite{mopac} using periodic boundary conditions to model de solid-sate materials \cite{solids}.  The software version parallelized for shared memory multiprocessors and graphic processor units \cite{maia} has been used.  Both the PM6 \cite{PM6} and PM7 \cite{PM7} Hamiltonians have been employed in the computations.  Accurate solid state calculations using MOPAC \cite{solids} require that a sphere with  \SI{8}{\angstrom} diameter can be fitted inside the computational domain.  This condition can be accomplished by extending the basic periodic unit used in the computations over several crystal unit cells.    The extent of the computational domain is defined in MOPAC using the keyword {\tt MERS=(n,m,l)} for $n$, $m$ and $l$ unit cells along the $a$, $b$, and $c$ crystal axes, respectively.  This procedure has obvious implications on the calculation of the normal modes of vibration, as discussed below.

For each material, the experimental crystal structure reported in the bibliography has been used as initial condition for the geometry optimization.  The calculations start with the thorough optimization of the geometry, followed by the analysis of the vibrational modes using the {\tt FORCE} keyword in MOPAC.  The absence of imaginary frequencies guarantees that the calculated geometries correspond to true energy minima.  Even though the newer PM7 parametrization generally provides better agreement between the predicted and experimental geometries, it is often found that PM6 provides better convergence behavior and it has been the preferred method in some of the cases in order to obtain the tightly optimized geometries required in the {\tt FORCE} calculations.  Since crystal lattice vibrations are particularly relevant to the terahertz spectra of materials, the crystal parameters of the optimized geometries computed with MOPAC are compared with the experimental values in order to assess the accuracy of the theoretical predictions, instead of using intra-molecular geometry parameters. 

The number of vibration modes calculated in a extended computational period is of the order of magnitude of the number of unit cells $n\times m\times l $ inside the computational domain times the number of actual k=0 vibrations of the material under study.  The modes of the extended computational cell include those sampled at locations different from $k=0$ in the first Brillouin zone associated to the fundamental unit cell of the material \cite{chamorro2}.  These additional vibrations will not show up in a theoretical study based on the crystal unit cell and they are not relevant for the estimation of the attenuation spectra of materials, since only $k\simeq 0$ transitions fulfill the momentum conservation considerations.  Even though, in some cases, the associated transition dipoles are significantly smaller than for any nearby $k\simeq 0$ transition \cite{chamorro2} and their inclusion does not really affect the interpretation of the results, this is not necessarily the case and these resonances should be eliminated from the analysis. Nevertheless, Amrhein \cite{amrhein} has shown that in an amorphous material the observed attenuation spectra corresponds to the phonon density of states. Therefore, for very large computational domains, the additional sampling of the phonon dispersion curves could be expected to reproduce the spectra of the amorphous phase.  

The Gabedit software package \cite{gabedit} has been used in the analysis of the results for the representation of the vibrations modes.

\section{Results and Discussion}

The vibration modes lying in the terahertz spectral region of two polymers: polyethylene and polyvinylidene fluoride (form II) have been first addressed.  The far-infrared properties of these two materials have been largely studied and the results obtained are compared with previous theoretical and experimental results.  In the terahertz band, contributions from inter-molecular interactions in polymers normally correspond to translatory and rotatory vibrations of the chains and  can be identified by their typical temperature dependence \cite{bershtein,wietzke2010}.  Intra-molecular terahertz vibrations include involve both localized torsions of functional groups and modes which are more delocalized along the chain \cite{bershtein}. 

 Next, we address the low-frequency vibrations of crystalline $\alpha$D-glucose.  The lowest lying vibrations of polycrystalline mono- and poly-saccharides determining their the THz spectra is dominated by lattice modes involving molecular hydrogen bond or/and van der Waals forces \cite{husain,walther,upadhya}.  Therefore, terahertz spectroscopy is particularly useful in the structural characterization of carbohydrates, highly relevant to biological functions.

\subsection{Polyethylene}

The polyethylene unit cell is orthorhombic with parameters shown in the first raw of table \ref{tablaPE} \cite{avitabile}. The space group is $Pnam$, with four CH\textsubscript{2} groups in the unit cell.

\begin{landscape}
\begin{table}
\centering
\begin{tabular}{l||ccccccccc}
\hline
&$a$ (\si{\angstrom})&$b$ (\si{\angstrom}) &$c$ (\si{\angstrom}) &{UME} (\si{\angstrom}) &$\alpha$&$\beta$&$\gamma$& UME&$\nu$(\si{\per\cm})\\
\hline\hline
Experimental             &$7.121$ & $4.851$  & $2.548$ &         &$90^o$ & $90^o$         & $90^o$    &          &$79.5$  \\\hline\hline
PM6 $2\times 2 \times 4$ & $7.842$ & $4.476$ & $2.536$ & $0.369$ & $105.85^o$ & $85.78^o$ & $92.84^o$ & $7.64^o$  & $68.01$\\
PM7 $2\times 2 \times 4$ & $7.102$ & $4.796$ & $2.526$ & $0.032$ & $89.89^o$ & $90.66^o$ & $90.38^o$ & $0.38^o$   & $83.70$\\\hline
PM6 $2\times 3\times 4$  & $7.843$ & $4.333$ & $2.536$ & $0.417$ & $95.74^o$ & $89.28^o$ & $89.68^o$ &  $2.26^o$  & $64.16$\\
PM7 $2\times 3\times 4$  & $7.095$ & $4.798$ & $2.524$ & $0.034$ & $ 89.95^o$ & $90.15^o$ & $89.86^o$ & $0.11^o$   & $88.67$\\\hline
PM6 $2\times 3\times 6$  & $7.129$ & $4.803$ & $2.537$ & $0.022$ & $90.03^o$ & $90.09^o$ & $89.91^o$ & $0.07^o$   & $81.69$\\
PM7 $2\times 3\times 6$  & $7.115$ & $4.774$ & $2.527$ & $0.035$ & $90.01^o$ & $89.63^o$ & $90.22^o$ & $0.20^o$   & $87.26$\\\hline
PM6 $3\times 4\times 9$  & $7.125$ & $4.811$ & $2.536$ & $0.019$ & $90.07^o$ & $90.07^o$ & $89.95^o$ & $0.06^o$  & $84.05$\\
PM7 $3\times 4\times 9$  & $7.066$ & $4.711$ & $2.527$ & $0.072$ & $90.04^o$ &  $90.06^o$ & $89.95$ & $0.05^o$   & $82.69$\\
\hline\hline
\end{tabular}
\caption{Experimental (low temperature) and computed data for the crystal structure parameters and the main vibration mode of crystalline polyethylene in the terahertz band.  Experimental data has been obtained from Ref.\cite{avitabile}.  The displayed values of unsigned mean error (UME) are in angstroms and degrees for lattice lengths and angles, respectively. }\label{tablaPE}
\end{table}
\end{landscape}

In the terahertz region, the spectrum of polyethylene (PE) is characterized by a main resonance at \SI{72}{\per\centi\metre} (\SI{2.2}{\tera\hertz}) at room temperature \cite{wietzke,tasumi}.   This absorption is originated from crystal lattice vibrations, and its frequency is found to vary both with the degree of crystallinity \cite{bank,dean}, and temperature \cite{bank,dean,birch}.  The resonance frequency  and is shifted to \SI{79.5}{\per\centi\metre} at low temperatures \cite{dean}.    Group theory predicts the existence of two active lattice modes,  B\textsubscript{1u} and B\textsubscript{2u}, associated with transverse displacements of the polymer chains and a forbidden transition  A\textsubscript{u}  with displacements along the polymer axis \cite{krimm,tasumi65,tasumi,bertie,frenzel}.  Even though the latter transition is forbidden in a perfect orthorhombic crystal arrangement \cite{krimm,tasumi65,tasumi}, the irregularities of glassy materials permit a relaxation of the selection rules and the observation of this vibration \cite{amrhein} as a faint attenuation band around \SI{60}{\per\centi\metre}.  In high-pressure annealed samples and at sufficiently  low temperatures (below \SI{170}{\kelvin}) an absorption band at \SI{39}{\per\centi\metre} has also been attributed to vibrations analogous to the  forbidden A\textsubscript{u} vibration of the orthorhombic PE that becomes active in the monoclinic form produced by the annealing \cite{frank}.  The second allowed mode has also been reported as a very low intensity absorption band in highly crystalline samples lying at \SI{107}{\per\cm} at liquid-nitrogen temperature \cite{fleming} and   \SI{109}{\per\centi\metre} at liquid helium temperature \cite{dean}.

\begin{figure}
\centering
\begin{tabular}{cc}
(a)&(b)\\
\includegraphics[width=6cm]{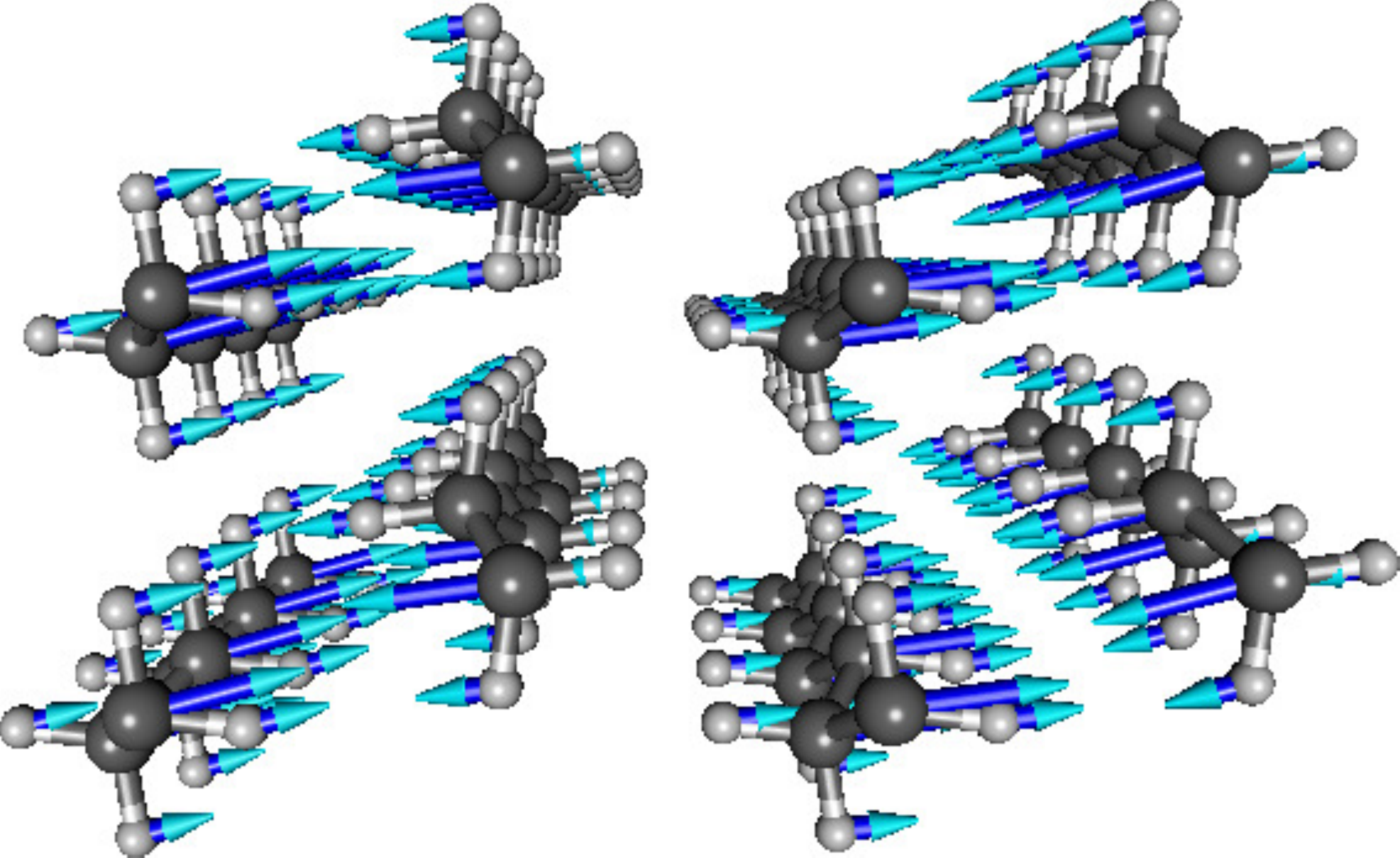}&\includegraphics[width=6cm]{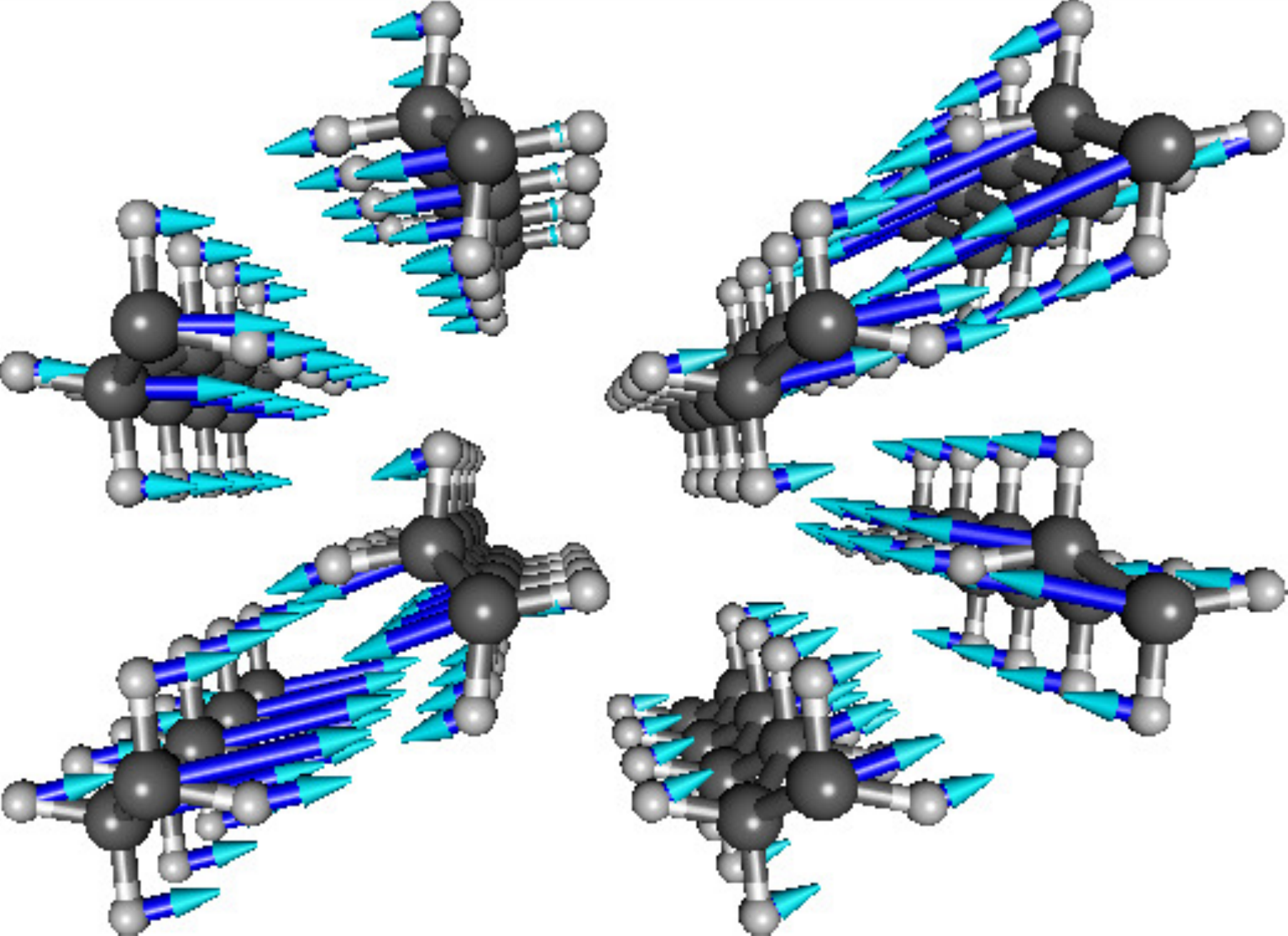}\\
(c)&(d)\\
\includegraphics[width=6cm]{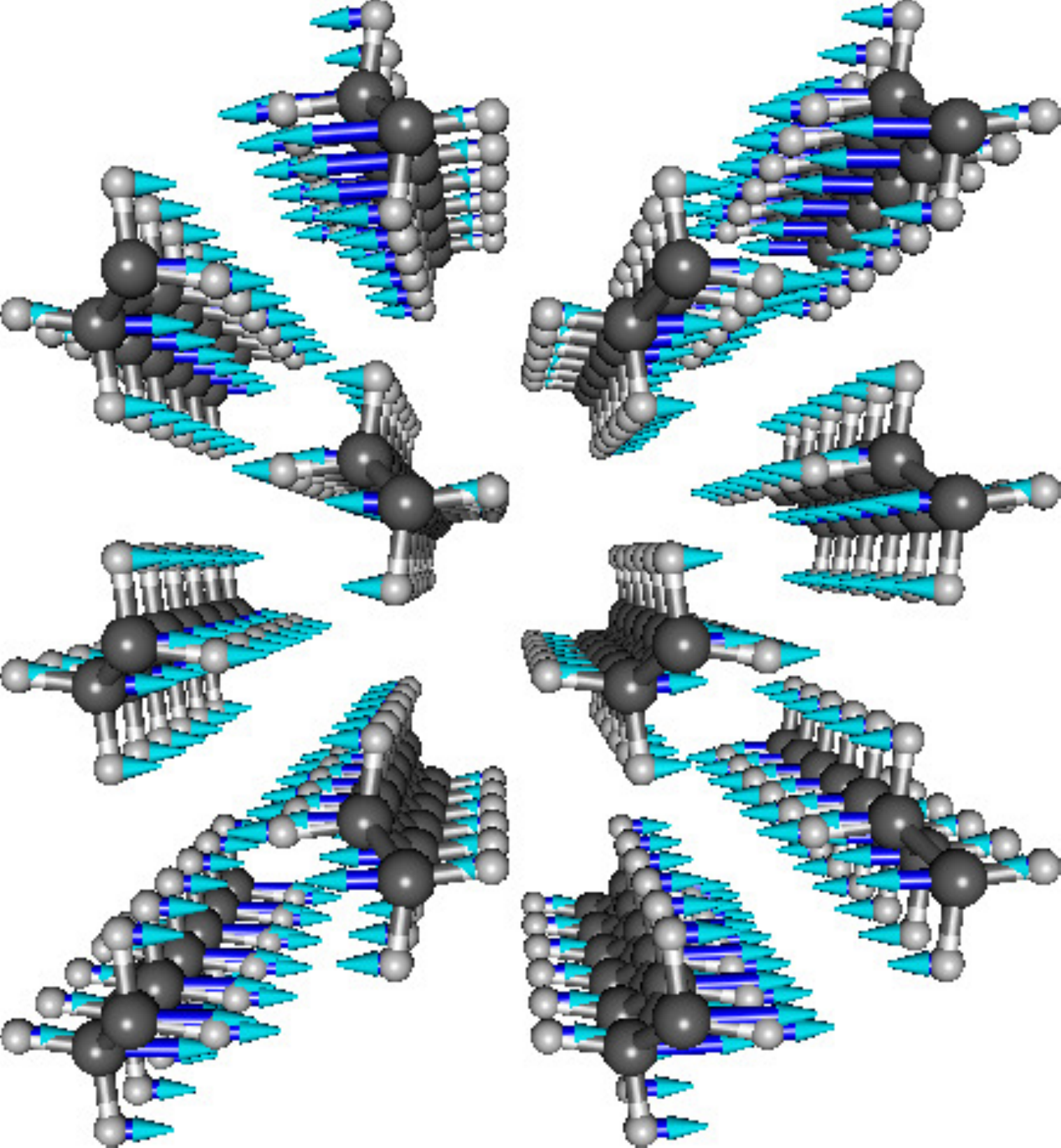}&\includegraphics[width=6cm]{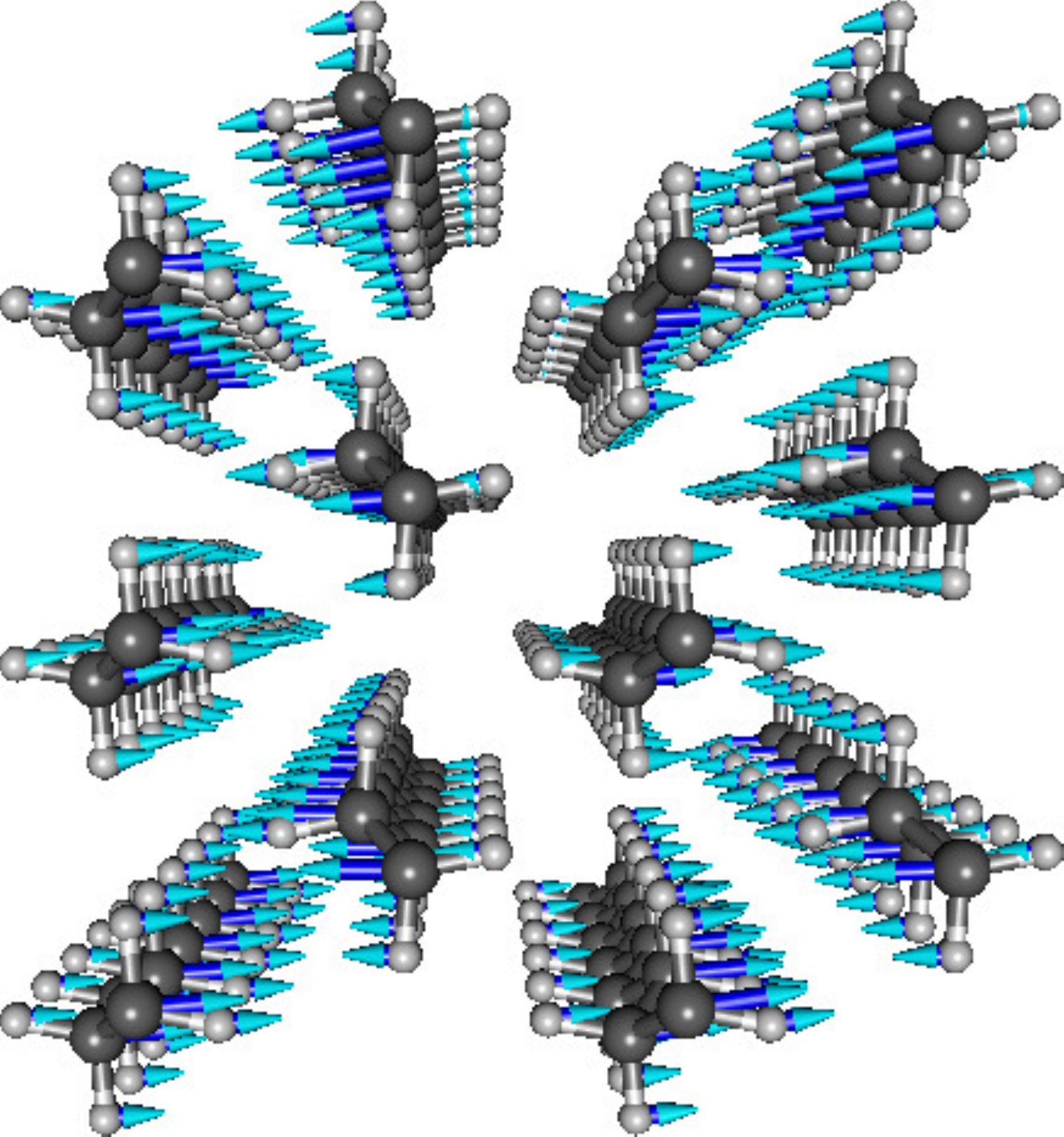}
\end{tabular}\caption{Main vibration mode in the terahertz spectral region for a $2\times 2 \times 4$ computational unit cell and the PM6 (a) and PM7 (b) Hamiltonians.  Results obtained using a period volume spanning $2\times 3 \times 6$  crystal cells and the PM6 and PM7 Hamiltonians are shown in subfigures (c) and (d), respectively.  The corresponding calculated frequencies are listed in table \ref{tablaPE}} \label{fig:vibracionesPE}
\end{figure}

The calculation of the vibration modes and transition dipoles using the PM6 and PM7 methods has been performed for different sizes of the computational domain.  In all cases, the calculations give an isolated or clearly dominant vibration mode in the vicinity of the the experimental value for the main PE absorption band at low temperatures.  This mode displays an in-phase oscillation in  all the crystal cells within the computational domain and, therefore, it corresponds to a $k=0$ vibration.  Also, the transverse displacements of the polymer chains result in an oscillating dipole moment oriented along the $a$ crystal axis, consistent with the observations in dichroism experiments \cite{bank}. Figure \ref{fig:vibracionesPE} illustrates this main vibration mode as obtained in four different computations with varying size of the computational domain and the Hamiltonian.  

Table \ref{tablaPE} shows the detailed comparison of the experimental results with those obtained using MOPAC as the size of computational domain and the Hamiltonian are varied.  The minimal computational cell that covers the \SI{8}{\angstrom} diameter sphere is a crystal volume of $2a\times 2b\times 4c$ crystal cells.  As discussed in the Methods section, this arrangement is obtained with the keyword  {\tt MERS=(2,2,4)}.  For this computational cell, the optimized geometry obtained with the PM7 is very close to the actual crystal, whereas the error for the calculations performed using the PM6 Hamiltonian are much larger, with a clearly distorted crystal geometry.  The numerical values corresponding to the wavenumber of the main vibration mode are also listed in the last column of table \ref{tablaPE}.  The accuracy obtained in the estimated values of the vibration frequencies are in accordance with the accuracies of the geometries.  For the minimal computational domain size, a $5.3\%$ deviation of the theoretical estimation with the experimental observation is obtained for the PM7 results whereas the less accurate PM6 geometry has an associated deviation of $14.5\%$ in the calculated frequency value.      

When the computational period is enlarged to $2\times 3\times 4$ crystal unit cells (the results are not shown) neither the geometries nor the frequencies show any improvement in their accuracy.  On the contrary, the results display larger deviations with the experimental values.  For a computational cell $2a\times 3b\times 4c$ the lateral dimensions along the $a$, $b$ and $c$ axes are very similar and approximately equal to \SI{14.2}{\angstrom}, \SI{14.6}{\angstrom} and \SI{15.3}{\angstrom} respectively.  In this case, the geometry predicted to a higher accuracy with the PM6 Hamiltonian and a similar improvement is observed in the estimation of the frequency, showing a relative deviation of $2.8\%$ with the experimental value. The displacements involved in the vibration mode shown in Figures \ref{fig:vibracionesPE} (c) and (d) also display a better match with the expected displacements along the $a$ direction when compared with the corresponding results for smaller computational volumes at (a) and (b), respectively. On the other hand, similar levels of accuracy are obtained when the computational unit cell is enlarged to the $3\times 4\times 9$ with approximate dimensions of \SI{21.4}{\angstrom}, \SI{19.4}{\angstrom} and \SI{22.9}{\angstrom} along the $a$, $b$ and $c$ axes, respectively.

\subsection{Polyvinylidene fluoride (form II)}

Polyvinylidene fluoride (PVDF) exists in different phases.  Form II (the $\alpha$-phase) has a monoclinic structure with space group $P2_1/c\left(C_{2h}^5\right)$ \cite{hasegawa}.   Both the intra and intermolecular vibrations of PVDF form II in the far infrared have been studied in \cite{rabolt,kobayashi}.  Two main absorption lines have been consistently identified at \SI{53}{\per\cm} and \SI{100}{\per\centi\metre} at room temperature \cite{rabolt}.  This results have been recently confirmed by THz-TDS measurements \cite{mori}, together with the identification of a third resonance at \SI{78.93}{\per\cm}.  The sensitivity to the crystallinity and working temperature permitted to identify  \cite{rabolt} the \SI{53}{\per\cm} line as a crystallinity lattice mode.  This line at  \SI{53}{\per\cm} shifts to \SI{60}{\per\cm} when the sample is cooled from \SI{300}{\kelvin} to \SI{90}{\kelvin} \cite{rabolt}.

Several theoretical studies have been performed using different quantum chemistry methods.   A semi-empirical CNDO method in \cite{correia} was used in \cite{correia} to optimize the geometry of a finite size chain of 20 monomers and to perform a vibrational analysis.  In \cite{ramer},  the crystal geometry and solid-state vibrations have been studied using a DFT method.  

The crystal structure of form II of PVDF used for initial condition in the geometry optimizations in this work was obtained from \cite{hasegawa}.  Experimental and theoretically predicted crystal parameters are shown in Table \ref{tablaPVDF}, together with the unsigned mean error (UME).  The minimal computational spatial period covers $2\times 1\times 2$ crystal unit cells.  The PM7 results show a significant deviation from the experimental values, but PM6 error in the lattice constants is comparable to that obtained with DFT calculations \cite{ramer}.  The PM7 method failed to converge for this computational cell size.  At larger sizes, no relevant improvement of the predictions of the crystal geometry is observed.

\begin{landscape}
\begin{table}
\centering
\begin{tabular}{l||cccccccc}
\hline
&$a$ (\si{\angstrom})&$b$ (\si{\angstrom}) &$c$ (\si{\angstrom}) &{UME} (\si{\angstrom}) &$\alpha$&$\beta$&$\gamma$& UME\\
\hline\hline
Experimental             &$4.96$ & $9.64$  & $4.62$ &         &$90^o$ & $90^o$         & $90^o$    &         \\\hline\hline
DFT                      & $5.18$  & $10.30$  & $4.70$ &  $0.320$  &           & $91^o$    &          &         \\\hline\hline
PM6 $2\times 1 \times 2$ & $5.467$ & $10.146$ & $4.727$ & $0.373$  & $88.29^o$ & $93.88^o$ & $91.38^o$ & $2.323^o$    \\\hline
PM6 $2\times 2\times 2$  & $5.489$ & $9.921$ & $4.738$ &  $0.309$ & $90.0^o$ & $93.07^o$ & $89.98^o$ &  $1.030^o$  \\
PM7 $2\times 2\times 2$  & $5.685$ & $9.953$ & $4.796$ &  $0.504$ & $ 88.38^o$ & $88.62^o$ & $86.21^o$ & $2.263^o$   \\\hline
PM6 $4\times 2\times 4$  & $5.463$ & $10.183$ & $4.726$ & $0.384$ & $90.24^o$ & $94.12^o$ & $89.83^o$ & $1.510^o$ \\
PM7 $4\times 2\times 4$  & $5.428$ & $10.357$ & $4.817$ & $0.461$ & $89.36^o$ & $90.68^o$ & $93.91^o$ & $1.743^o$  \\
\hline\hline
\end{tabular}
\caption{Experimental (low temperature) and computed data for the crystal structure parameters in the terahertz band.  Experimental data has been obtained from Ref.\cite{hasegawa} and DFT from Ref. \cite{ramer}. }\label{tablaPVDF}
\end{table}
\end{landscape}

\begin{figure}
\centering
\includegraphics[width=12cm]{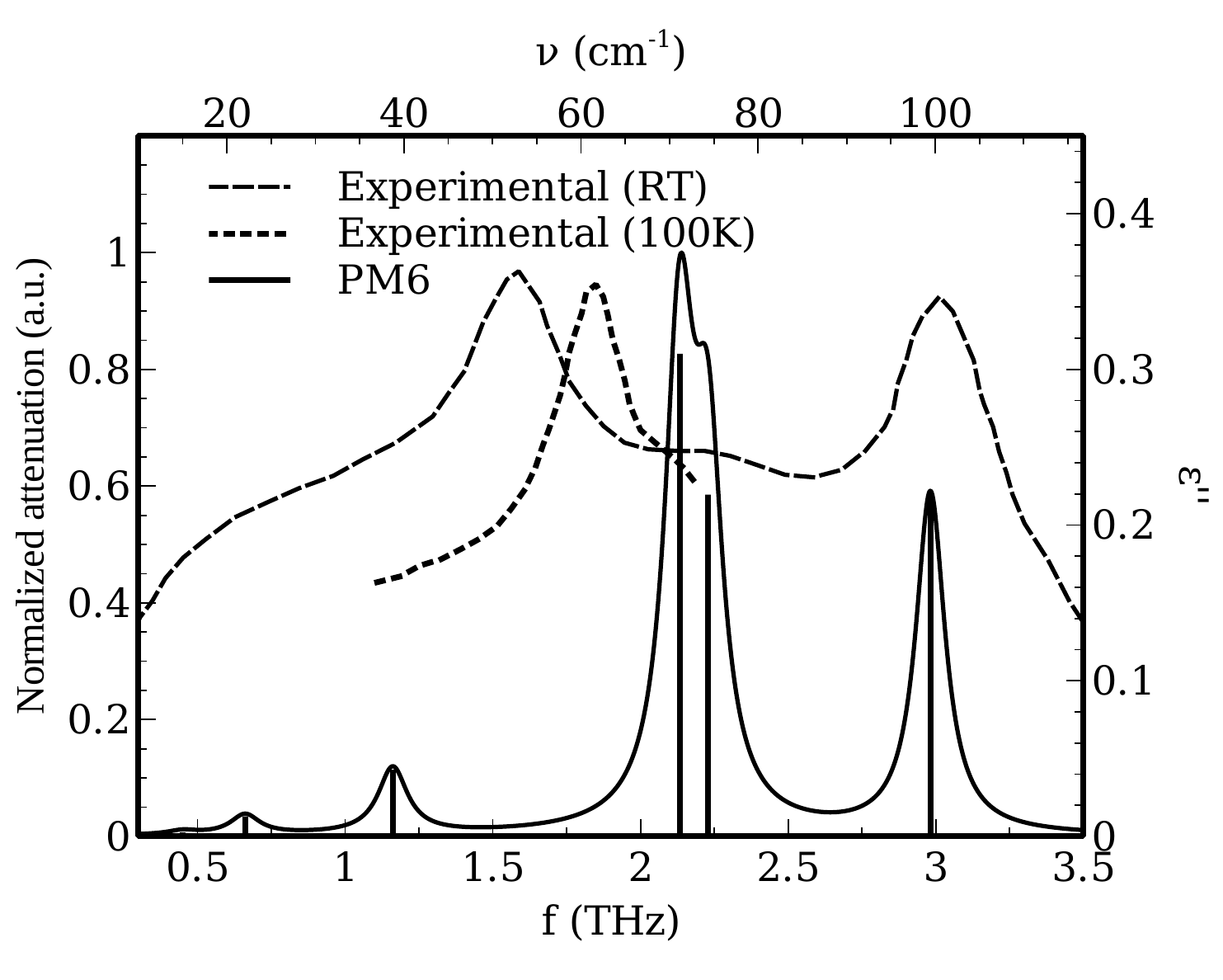}
\caption{The room temperature terahertz spectrum of PVDF form II \cite{mori} is plotted with dashed lines and the lowest resonance peak at \SI{100}{\kelvin} \cite{mori} with dotted lines.  Experimental data corresponds to the imaginary part of the relative dielectric permittivity.  The calculated vibration frequencies and normalized intensities in the spectral range up to  \SI{3.5}{\tera\hertz} are shown with solid lines.  The numerical results have been obtained using MOPAC with the PM6 Hamiltonian and periodic boundary conditions with a computational period comprising $2\times 1\times 2$ crystal unit cells.  The shape obtained by convolving the spectral lines with a Lorentzian function of an arbitrary width is also plotted with a solid line.} \label{fig:espectroPVDF}
\end{figure}

\begin{figure}
\centering
\begin{tabular}{cc}
(a)&(b)\\
\includegraphics[width=6cm]{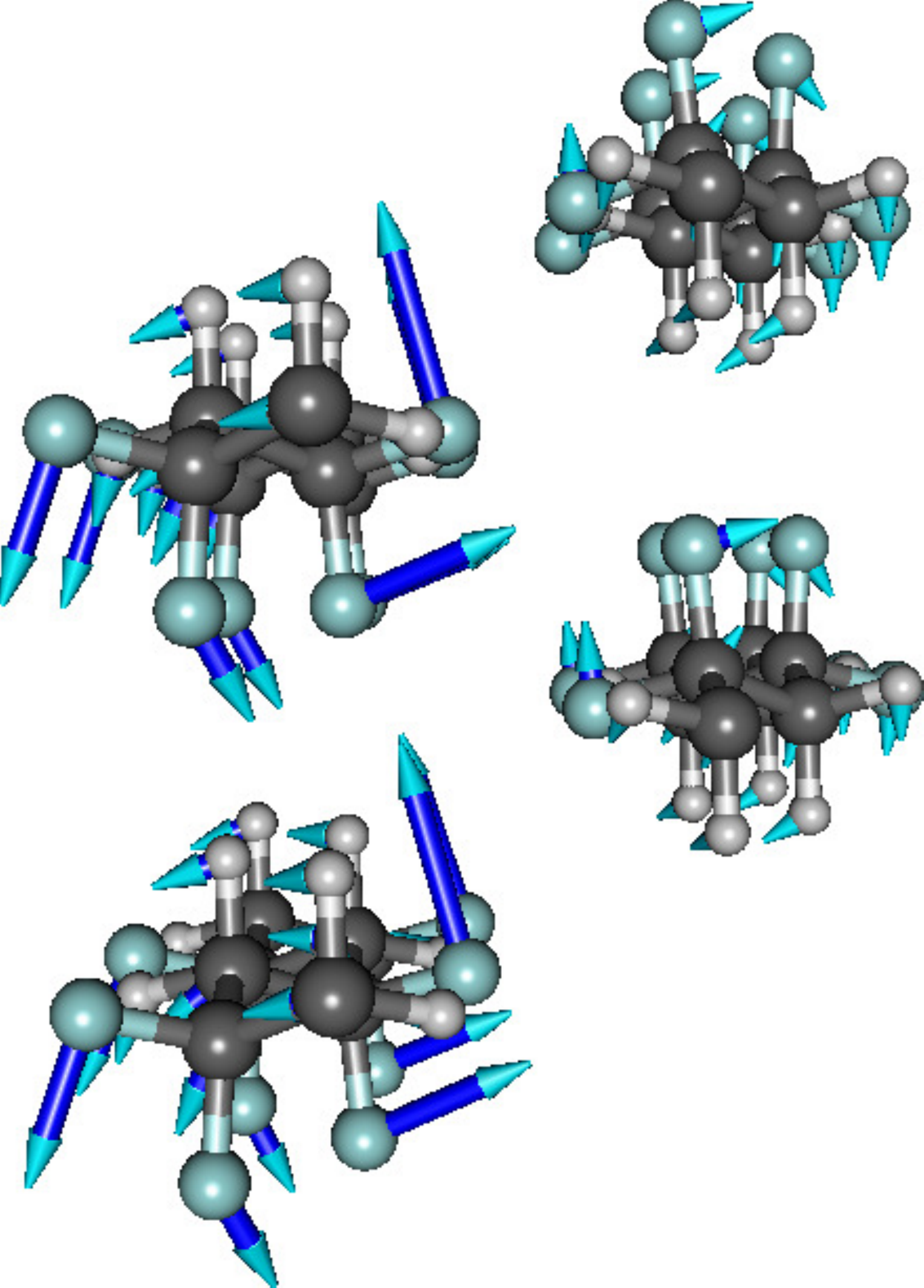}&\includegraphics[width=6cm]{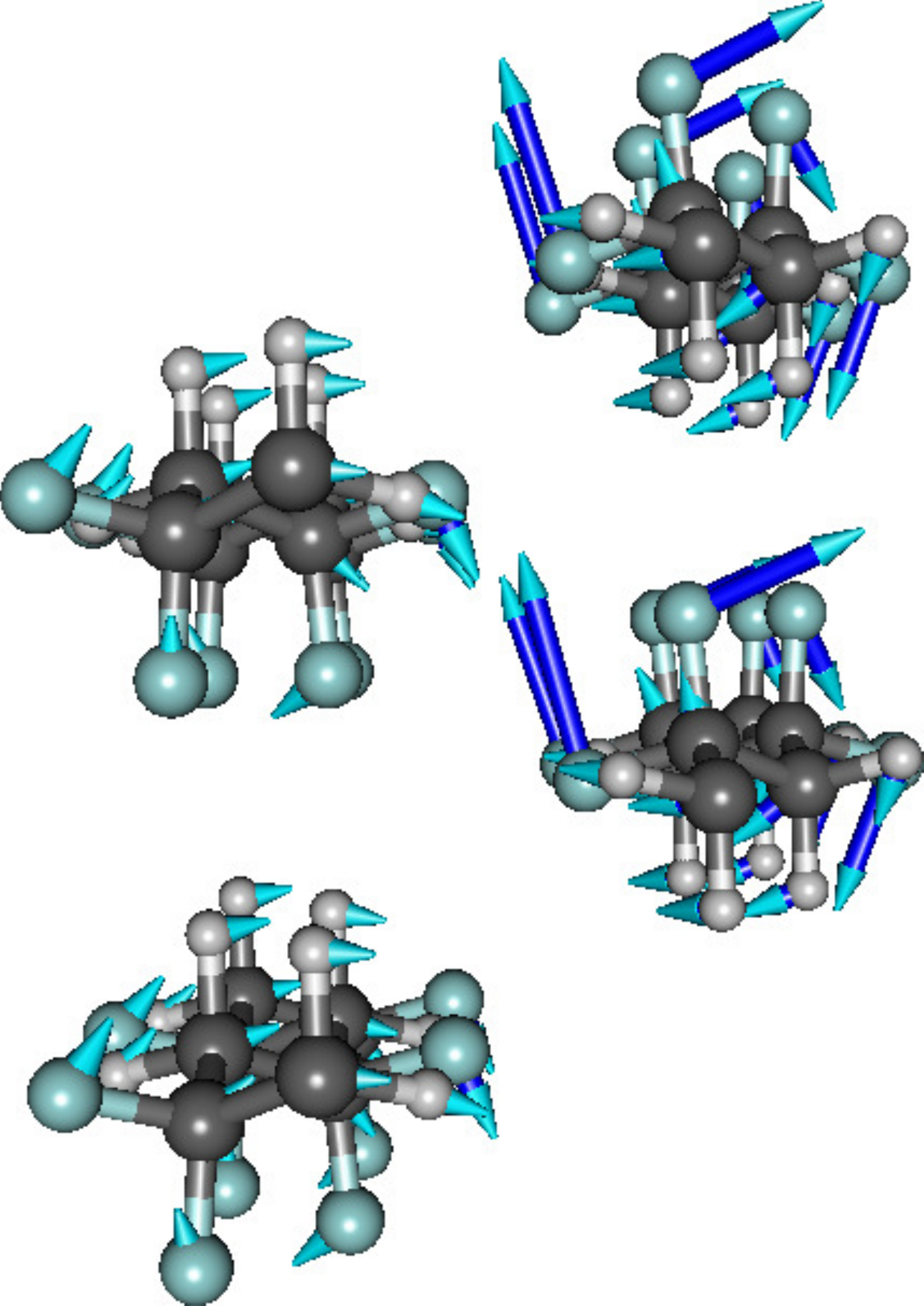}
\end{tabular}
\caption{Normal vibration modes of PVDF calculated for the $2\times 1\times 2$ computational cell with the PM6 method lying at \SI{71.14}{\per\cm} (a) and \SI{74.36}{\per\cm} (b) analogues to the $R_c^\pi$ and $R_c^o$ lattice vibrations of the actual crystal \cite{kobayashi}, respectively.} \label{fig:librations}
\end{figure}
. 

The experimental spectrum of PVDF is shown in Figure \ref{fig:espectroPVDF} together with the results calculated with MOPAC using the PM6 method.  The experimental resonance at \SI{100}{\per\cm} is calculated at \SI{99.46}{\per\cm}, with a relative error of $0.54\%$.  Previous studies using force constant analysis \cite{kobayashi} and DFT \cite{ramer} have predicted its value at \SI{94}{\per\cm} and \SI{105}{\per\cm}, with relative errors of $6\%$ and $5\%$, respectively, within this assignment.  

The mode calculated at \SI{71.14}{\per\cm} is depicted in Figure \ref{fig:librations} (a) and  corresponds to a lattice libration with $\pi$ phase difference between the two chains in the unit cell, similar to the $R_c^\pi$ mode described in \cite{kobayashi}.  Therefore, it can be assigned to the experimental \SI{60}{\per\cm} resonance, with a relative error of $19\%$.  DFT calculations \cite{ramer} predict this mode at \SI{55}{\per\cm}, with a relative error of $8.3\%$.  An expected additional blue-shift of the resonance at lower temperatures would tend to equalize the relative errors in the PM6 and DFT methods.

The resonance calculated at \SI{74.36}{\per\cm} could be assigned to the absorption measured at \SI{78.93}{\per\cm} in Ref. \cite{mori}. The normal mode is depicted in Fig. \ref{fig:librations} (b) and is similar to the $R_c^0$ libration in the experimental crystal arrangement.  Even though the corresponding transition in the perfect crystal is not optically active, the distortion of the computational geometry can relax the selection rules.  Similarly, a non-perfect degree of crystallinity in the sample can allow for the experimental observation of this absorption, as commented previously in the case of PE.  In the calculated resonances displayed in Fig. \ref{fig:librations}, the amplitude of the displacements of the fluorine atoms at each of the two chains in the unit cell alternatively dominates in the two modes.

 The three resonances calculated at \SI{14.99}{\per\cm}, \SI{22.05}{\per\cm} and \SI{38.73}{\per\cm} with very small transition dipoles correspond to translations of the polymer chains with a $\pi$ dephasing between the two chains in the unit cell.

\subsection{$\alpha$-D-glucose}

$\alpha$-D-glucose crystallizes in the orthorombic system with $a=\SI{10.3662(9)}{\angstrom}$, $b=\SI{14.8506(16)}{\angstrom}$ and $c=\SI{4.9753(3)}{\angstrom}$ \cite{brown}.  There are four molecules in the elementary crystalline cell, which are not hydrogen bonded.  The hydrogen bonds are formed with the molecules of neighboring cells.  

Experimental studies of the FIR spectrum of glucose were performed in the 1970s using  a grating spectrometer \cite{hineno72,hineno76},  Fourier Transform spectroscopy \cite{husain} in the 1980s and, more recently, using terahertz time-domain spectroscopy \cite{walther,upadhya,zheng}.  Theoretical calculations of the solid-state lattice vibration \cite{dauchez} have also been performed, permitting to determine a force field for the $\alpha$ and $\beta$ anomers of glucose.  

The solid-state geometry was optimized using the PM6 method with periodic boundary conditions and a computational domain spanning two unit cells in the $c$ direction: {\tt MERS=(1,1,2)}. The structure used as the initial condition was taken from \cite{brown}. The crystal parameters of the calculated solid state geometry are $a=\SI{10.263}{\angstrom}$, $b=\SI{15.228}{\angstrom}$, $c=\SI{4.080}{\angstrom}$, $\alpha=90.01^o$, $\beta=89.99^o$ and $\gamma=90.00^o$, in good agreement with the experimental values.  The UME associated to the computed lattice constants is \SI{0.458}{\angstrom}.  In the calculation of the vibration modes, using the {\tt FORCE} keyword in MOPAC, the {\tt PRECISE} keyword was also employed.  The use of this keyword, which permits to eliminate the quartic contamination, is not encouraged in the MOPAC documentation since, in general, does not have a major impact.  In this case, however, a correction of $8.7\%$ in the frequency of the dominant mode in the range from  \SIrange{0}{2}{\tera\hertz} has been found, providing a better agreement with the experimental results.  In this spectral region, $\alpha$-D-glucose has a dominant peak at \SI{1.46}{\tera\hertz} (\SI{48.7}{\per\cm}) at \SI{4}{\kelvin} \cite{upadhya} that experiences a very small red shift at room temperature.

\begin{figure}
\centering
\includegraphics[width=12cm]{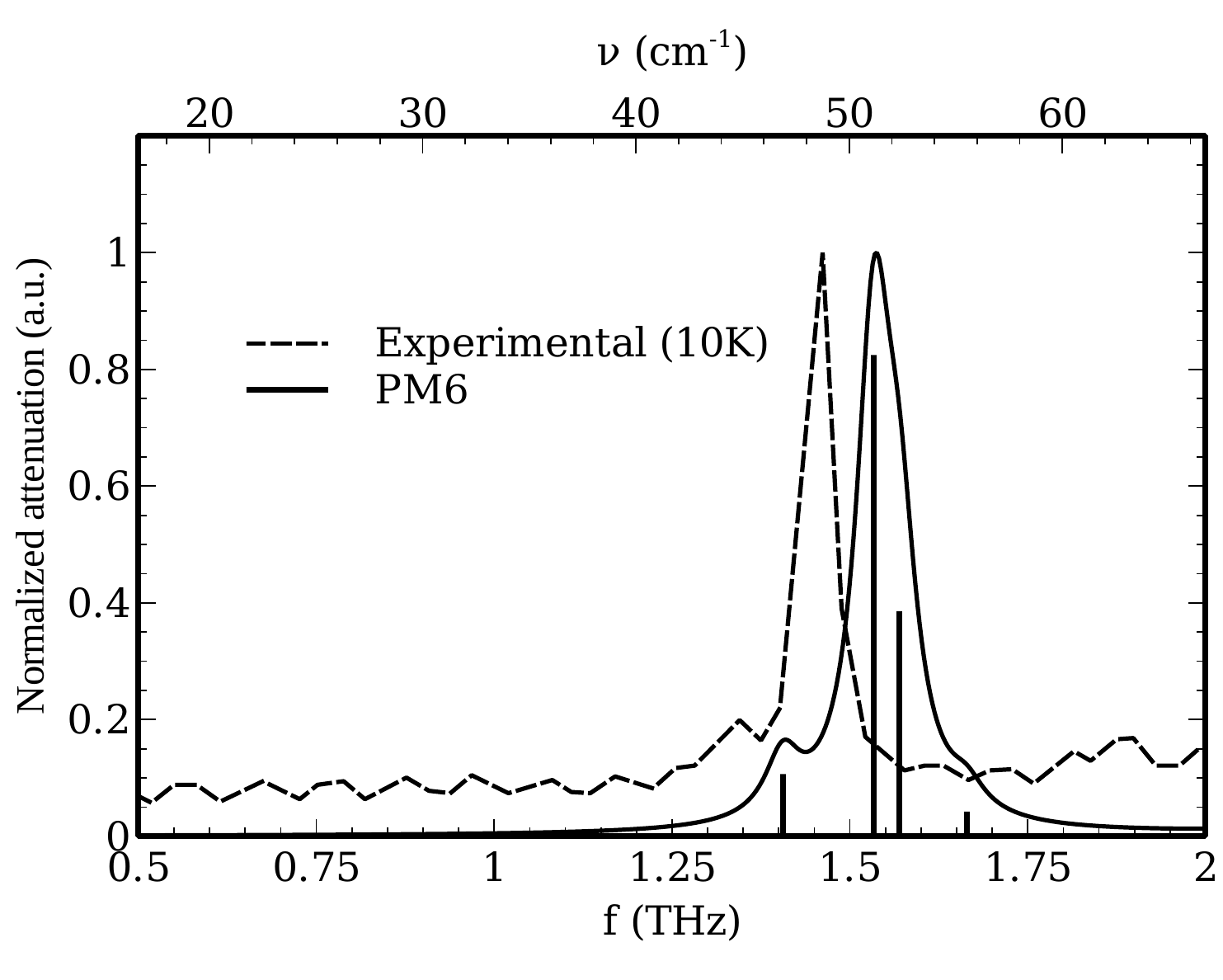}
\caption{Experimental terahertz attenuation of $\alpha$-D-glucose at low temperature from \cite{walther} and vibration frequencies and normalized intensities in the spectral range between \SI{20}{\per\centi\metre} and \SI{130}{\per\centi\metre} predicted using MOPAC with the PM6 Hamiltonian for the minimal computational cell comprising $2\times 1\times 2$ crystal unit cells.  } \label{fig:espectroglucosa}

\end{figure}

The experimentally measured spectrum at \SI{10}{\kelvin} is compared with the calculated results in figure \ref{fig:espectroglucosa} where only vibrations with in-phase oscillations in the two the crystal cells within the computational domain relevant for the terahertz absorption have been considered.  In this case, the description of the displacements of the lattice vibrations is rather involved \cite{dauchez}.   MOPAC results provide a good qualitative description of the response, even though the frequencies are blue-shifted by about \SI{4}{\per\cm}, with a relative error of about $8\%$.

\section{Conclusion}

The performance of two semi-empirical NDDO methods (PM6 and PM7) for the prediction of low frequency vibrations in the crystalline solid state has been addressed.  The study has been based on the terahertz properties of three different largely studied materials: PE, PVDF and $\alpha$-D-glucose.

The PM7 parametrization includes several improvements over PM6 and better accuracy in the computed geometries and, therefore, in the vibration modes is expected.  Nevertheless, in two of the three cases studied (PVDF and glucose) the convergence properties of the PM6 method where better, providing the required highly optimized geometry for the vibrations calculations.  The accuracy in the solutions when the size of the computational domain was increased was also studied.  Only in the case of PE with the PM6 Hamiltonian an improvement was observed.  In all the other cases, such enlargement of the computational domain had no effect on the accuracy of the predicted geometries.     

Solid-state computations using MOPAC require that the computational domain exceeds a minimum size.  Therefore, it is often necessary to perform the calculations over an extended domain.  This has the drawback that the vibration modes with all the crystal cells oscillating in phase relevant for the far-infrared spectra have to be selected from the whole set of calculated modes.

The results obained in this work show a good qualitative description of the terahertz spectra of the materials studied.  In the case of PE, the main vibration mode is predicted with a relative error of $2.8 \%$.  For PVDF, the peak at \SI{100}{\per\cm} is predicted with a very small error of $0.54\%$.  Nevertheless, the error in the calculated frequency for the lowest vibration mode is $19 \%$.  The study based on the PM6 methods permits to assign the third resonance in this band found in Ref. \cite{mori}. For glucose, the error in the prediction of the main resonance in the band from \SIrange{0}{2}{\tera\hertz} is approximately $8\%$.

\end{document}